\documentclass[12pt]{iopart}

\def\newblock{\hskip .11em plus .33em minus .07em}

\usepackage{graphicx}
\usepackage{dcolumn}
\usepackage{bm}
\usepackage{bbm}
\usepackage{epsfig}
\usepackage{mathrsfs}
\usepackage{stmaryrd}
\usepackage{color}
\usepackage{dsfont}
\usepackage{iopams}
\usepackage{psfrag}

\newcommand{\bra}[1]{\langle\,{#1}\, |}
\newcommand{\ket}[1]{|\,{#1}\,\rangle}

\newcommand{\Real}{\mbox{Re}}

\newcommand{\ElTransE}{ \varepsilon}

\newcommand{\LichtFr} { \Omega}  
  \newcommand{\LichtPol} {\hat{\mathcal{E}}}  
\newcommand{\V}{V}
\newcommand{\CrossSec}{\sigma}

\newcommand{ \dip}{ \vec{\mu}}           

\newcommand{ \VibFr} { \omega}                       

\newcommand{ \vc} { \kappa}                       

\newcommand{\OOp}{D}
\newcommand{\OOpBar}{\bar{\OOp}}

\newcommand{\HamAggEl}{H_{\rm el}}

\newcommand{\HamAggBath}{H_{\rm env}}
\newcommand{\HamAggInt}{H_{\rm int}}

\newcommand{\AnzMon}{N}

\newcommand{\aDestroy}{a}
\newcommand{\aPlus}{a^{\dagger}}

\newcommand{\BathCor}{\alpha}

\newcommand{\kB}{k_B}

\newlength{\mylenunit} 

\newcommand{\refcite}[1]{\cite{#1}}

\begin{document}


\title[Spectral properties of molecular oligomers]
{Spectral properties of molecular oligomers.\\ A non-Markovian quantum state diffusion approach.}
\author{J.~Roden$^1$, W.T.~Strunz$^2$ and A.~Eisfeld$^1$}
\address{$^1$Max Planck Institute for the Physics of Complex Systems, N\"othnitzer
  Strasse 38, 01187 Dresden, Germany,\\
$^2$Institut f\"{u}r Theoretische Physik,
Technische Universit\"at Dresden, D-01062 Dresden, Germany }
\ead{eisfeld@pks.mpg.de}


\begin{abstract}
Absorption spectra of small molecular aggregates (oligomers) are considered.
The dipole-dipole interaction between the monomers leads to shifts of the oligomer spectra with respect to the monomer absorption.
The line-shapes of monomer as well as oligomer absorption depend strongly on the coupling to vibrational modes. 
Using a recently developed approach [Roden et.~al, PRL 103, 058301] we investigate the length dependence of spectra of  one-dimensional aggregates for various values of the interaction strength between the monomers.
It is demonstrated, that the present approach is well suited to describe the occurrence of the J- and H-bands.
\end{abstract}

%
\section[Introduction]{Introduction\label{introduction}}
Due to their unique optical and energy-transfer properties molecular
aggregates have attracted researchers for decades \cite{FrTe38_861_,Ko96__,EiKnKi09_658_}.
In recent years there has been a growing interest in one-dimensional molecular
aggregates consisting of a small number $N$ (ranging from only two to a few
tens) of molecules \cite{KoHaKa81_498_,SeWiRe09_13475_,RoEiBr08_258_,WeSt03_125201_,SuGoRe08_877_,Sc96__,GuZuCh08_2094_}, where finite size effects play an important role. 
Such systems are e.g.\ dendrimers \cite{SuGoRe08_877_}, stacks of organic dye
molecules \cite{SeWiRe09_13475_} or the light harvesting complexes in
photosynthesis \cite{GrNo06_793_,ReChAs09_184102_,ReMaKue01_137_}.
Usually in these aggregates there is negligible overlap of electronic wavefunctions between neighboring monomers. However, there is often a strong (transition) dipole-dipole interaction, that leads to eigenstates where an electronic excitation is coherently delocalized over many molecules of the aggregate, accompanied by often drastic changes in the aggregate absorption spectrum with respect to the monomer spectrum.
Since these changes depend crucially on the conformation of the aggregate, commonly the first step in the investigation of such aggregates 
 is optical spectroscopy.
Important properties, like the strength of the dipole-dipole interaction can
often already be deduced from  these  measurements \cite{KoHaKa81_498_,Sc96__,EiBr06_376_,FiKnWi91_7880_}.
 For small aggregates, often referred to as oligomers, the finite size leads to pronounced effects in their optical properties.

The interpretation of  measured absorption spectra is complicated by the fact
that the electronic excitation couples strongly to distinct vibrational modes
of the molecules and to the environment, leading to a broad asymmetric monomer
absorption spectrum that  often  also shows a vibrational progression \cite{KoHaKa81_498_,EiBr06_376_}.
Because of this (frequency dependent) interaction with a quasi-continuum of vibrations,  the theoretical treatment, even of quite small oligomers (of the order of five to ten monomers), becomes a challenge. 
Exact diagonalization \cite{FuGo64_2280_,SeWiRe09_13475_,RoEiBr08_258_,BoTrBa99_1633_,Sp02_5877_} is restricted to a small (one to five) number of vibrational modes. 
Furthermore, to compare with experiments, the resulting ``stick-spectra'' have to be convoluted with some line-shape function to account for the neglected vibrations.
For large aggregates the ``coherent exciton scattering'' (CES) approximation,
which works directly with the monomer absorption line-shape as basic
ingredient \cite{BrHe70_1663_}, has been shown to give very good agreement
with experiment \cite{EiBr06_376_,EiBr07_354_,EiKnBr07_104904_}. 
However, one expects  problems for small ($N<10$) oligomers \cite{RoEiBr08_258_}.

To treat oligomers with strong coupling to vibrational modes, recently an
efficient approach was put forward \cite{RoEiWo09_058301_} which is based on a non-Markovian stochastic
Schr\"odinger equation.
In the present paper this approach  will be used to investigate the absorption of linear one-dimensional oligomers.

The paper is organized as follows:
In section \ref{sec:model} the Hamiltonian of the aggregate is  introduced followed by a brief review of  our method to calculate the absorption spectrum in section \ref{sec:method}. 
The calculated oligomer spectra are presented and discussed in section \ref{sec:results}. 
We conclude in section \ref{sec:conclusions} with a summary and an outlook.

\section{The model Hamiltonian}
\label{sec:model}
In this work we will use a commonly applied model  of a molecular aggregate
including coupling of electronic excitation to vibrations \cite{Sc96__,MaKue00__,AmVaGr00__}.
We consider an oligomer consisting of $N$  monomers.  The transition energy between the electronic ground and electronically excited state of monomer $n$ is denoted by $\ElTransE_n$.
We will investigate absorption from the state where all monomers are in their ground state.
A state in which monomer $n$ is  excited and all other 
monomers are in their ground state is denoted by $\ket{\pi_n}$. 
States in which the whole aggregate has more than one electronic excitation are not considered.
 We expand the total Hamiltonian  
\begin{equation}
\label{eq:ham_tot}
H=\HamAggEl+\HamAggInt+\HamAggBath
\end{equation}
 of the interacting monomers and the vibrational environment with respect to the electronic ``one-exciton'' states $\ket{\pi_n}$.  
The purely electronic Hamiltonian
\begin{equation}
\label{eq:Ham_sys}
\HamAggEl=\sum_{n,m=1}^N\Big(\ElTransE_n\delta_{nm}+ 
\V_{nm}\Big)\ket{\pi_n}\bra{\pi_m}
\end{equation}
contains the interaction $V_{nm}$ which describes exchange of excitation between monomer $n$ and $m$.
The environment of each monomer is taken to be a set of harmonic modes
described by the environmental Hamiltonian  
\begin{equation}
\HamAggBath=\sum_{n=1}^{\AnzMon}\sum_j \hbar \VibFr_{nj} \aPlus_{nj}  
\aDestroy_{nj}.
\end{equation}
Here $\aDestroy_{nj}$ denotes the annihilation operator of 
mode $j$ of monomer $n$ with frequency $\VibFr_{nj}$.
The coupling of a local electronic  excitation to its vibrational environment is assumed to be linear and the corresponding interaction Hamiltonian is given by
\begin{equation}
\HamAggInt=-\sum_{n=1}^{\AnzMon} \sum_j \vc_{nj} (\aPlus_{nj}  
+\aDestroy_{nj})\ket{\pi_n}\bra{\pi_n},
\end{equation}
where the coupling constant $\vc_{nj}$ scales the coupling of the excitation on monomer $n$ to the mode $j$ with frequency $\omega_{nj}$ of the local vibrational modes.
This interaction is conveniently described by  the spectral density
\begin{equation}
J_n(\omega)=\sum_j |\vc_{nj}|^2 \delta (\omega-\omega_{nj})
\end{equation}
of monomer $n$.
It turns out that the influence of the environment  of monomer $n$ is encoded
in the bath correlation function \cite{MaKue00__}, which at temperature $T$ reads
\begin{equation}
\BathCor_n(\tau)= \int d\omega\ J_n(\omega)\Big( \cos(\omega \tau)
\coth\frac{\hbar \omega}{2 \kB T} - i \sin(\omega \tau) \Big). 
\end{equation}
In this work we restrict the discussion to the case $T=0$ in which the bath-correlation function $\BathCor_{n}(\tau)$ reduces to
\begin{equation}
\label{eq:BathCor}
\BathCor_{n}(\tau)= \int d\omega\ J_n(\omega) e^{-i \omega \tau}= \sum_j |\vc_{nj}|^2 e^{-i \omega_{nj} \tau}.
\end{equation}
When the environment has no ``memory'', i.e.\ $\BathCor_{n}(\tau) \propto \delta(\tau)$ it is termed Markovian, otherwise it is non-Markovian.

\section{Method of calculation}
\label{sec:method}
We use a recently developed method to describe the interaction of the
excitonic system with a non-Markovian environment \cite{RoEiWo09_058301_} which is based on ideas from a stochastic Schr\"odinger equation approach to open 
quantum system dynamics \cite{DiSt97_569_,DiGiSt98_1699_}. 
Within this approach it is possible to exactly describe the dynamics governed by a Hamiltonian of the form of Eq.~(\ref{eq:ham_tot}) by a stochastic Schr\"odinger equation in the Hilbert space of the electronic system alone. 
However, due to the appearance of  functional derivatives in this stochastic Schr\"odinger equation, the general method is of limited practical use.
To circumvent this problem in Ref.~\refcite{RoEiWo09_058301_} these functional derivatives were approximated by operators $\OOp^{(m)}(t,s)$ which are independent of the stochastics and can be obtained from a set of coupled differential equations. 
   
Within this approximation the cross-section for absorption of light with 
frequency $\LichtFr$  at zero-temperature is simply given by
\begin{equation}
\label{app:CrossSec:c_nm}
\CrossSec(\LichtFr)=\frac{4 \pi}{\hbar c} \LichtFr\ 
\Real \int_{0}^{\infty}\!dt e^{i  \LichtFr t}
\langle\psi(0)|\psi(t)\rangle
\end{equation}
 with initial condition
\begin{equation}
\label{eq:initial_state}
|\psi(0)\rangle=\sum_{n=1}^N (\LichtPol\cdot \dip_n)\ket{\pi_n},
\end{equation}
where the orientation of the monomers enters  via the transition 
dipoles $\dip_n$ and the polarization of the light $\LichtPol$.
The state $|\psi(t)\rangle$ is obtained from solving a Schr\"odinger equation in the small Hilbert space of the excitonic system
\begin{eqnarray}
\label{Schr_final}
\partial_t |\psi(t)\rangle
  & = &
-\frac{i}{\hbar} \HamAggEl  |\psi(t)\rangle  +\sum_{m} |\pi_m\rangle\langle\pi_m|
\OOpBar^{(m)}(t)|\psi(t)\rangle
\end{eqnarray}
 with initial condition (\ref{eq:initial_state}).
The operator $\OOpBar^{(m)}(t)$ appearing on the right hand side of equation
(\ref{Schr_final}) contains the interaction with the environment upon
excitation of monomer $m$ in an approximate way \cite{RoEiWo09_058301_}, and is given by  
\begin{equation}
\label{def:OOpBar0}
\OOpBar^{(m)}(t)=\int_0^t\! d{s}\, \BathCor_{m}(t-s) \OOp^{(m)}(t,s),
\end{equation}
where  $\BathCor_{m}(t-s)$ is defined in Eq.~(\ref{eq:BathCor})
and the operator $\OOp^{(m)}(t,s)$ is obtained by solving
\begin{eqnarray}
\label{oop}
\partial_t\OOp^{(m)}(t,s)
&=&\Big[-\frac{i}{\hbar}\HamAggEl, \OOp^{(m)}(t,s)  \Big]\\
&&+\sum_l\Big[ \ket{\pi_{l}}\bra{\pi_{l}}
\nonumber
  \OOpBar^{(l)}(t) ,  \OOp^{(m)}(t,s)\Big],
\end{eqnarray}
with initial condition 
$\OOp^{(m)}(t=s,s)=-|\pi_m\rangle\langle\pi_m|$.

This method allows a numerically fast and simple treatment of an assembly of
interacting monomers with a complex environment.

\section{Results}
\label{sec:results}

 \begin{figure}[btp]
\psfrag{M1}{$\vec{\mu}_1$}
\psfrag{M2}{$\vec{\mu}_2$}
\psfrag{M3}{$\vec{\mu}_3$}
\psfrag{M4}{$\vec{\mu}_4$}
\psfrag{M5}{$\vec{\mu}_5$}
\centering
 \centerline{\psfig{file=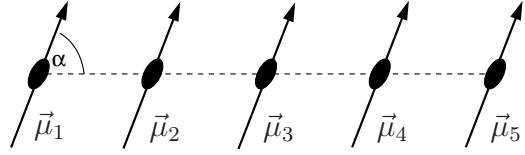,height=0.8in,width=2.7in}}
 \vspace*{8pt}
 \caption{\label{fig:skizze} Sketch of the geometry  considered for the case $N=5$. The points mark the positions of the monomers and the arrows indicate the direction of the respective transition dipoles. }
 \end{figure}
Using the method described in the previous section we will now investigate the dependence of oligomer absorption spectra on the number $N$ of monomers and the interaction strength between them.
To this end we choose a one dimensional arrangement of the monomers and take
 the  transition energies to be equal, $\ElTransE_{n}=\ElTransE$.  We  take only coupling between nearest neighbors into account, which we assume to be equal for all neighboring monomers, i.e.\ $V_{n,n+1}=V$ for all $n$. 
For simplicity we also take all transition dipole moments $\dip_n$ to be
identical and parallel to the polarization of the light. 
This arrangement is sketched in Fig.~\ref{fig:skizze}.
 \begin{figure}[bt]
\psfrag{energ2}{\scriptsize $\hbar\omega$ $[{\rm eV}]$}
\psfrag{specdens}{\hspace{-0.0\mylenunit}\scriptsize $J(\omega)$}
\psfrag{energy}{\hspace{-0.08\mylenunit}\raisebox{-0.\mylenunit}{\footnotesize  Energy $[{\rm eV}]$}}
\psfrag{absorption}{\footnotesize \hspace{-0.09\mylenunit}Absorption [arb. u.]}
 \centerline{\psfig{file=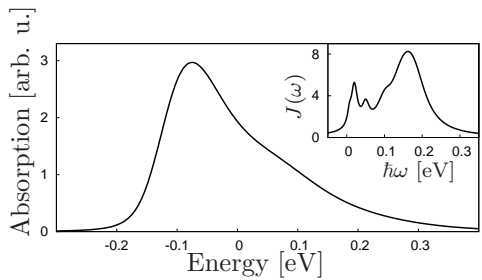,width=.8\mylenunit}
 }
 \vspace*{8pt}
 \caption{\label{fig:monomer} Absorption spectrum of a single monomer taken in the calculations. The inset shows the corresponding spectral density (unit $10^{-2}\ \rm{eV}/\hbar$). }
 \end{figure}
Before including the  complexity of the vibrations we will briefly discuss the purely electronic case, given by Eq.~\ref{eq:Ham_sys}. 
For the arrangement described above the Hamiltonian  Eq.~\ref{eq:Ham_sys} can be diagonalized analytically to obtain the eigenenergies $E_j=\ElTransE+2 V \cos(\pi j/(N+1))$ where the sign and magnitude of $V$ depend crucially on the angle $\alpha$ between the transition dipoles and the axis of the oligomer. 
The oscillator strength $F_{j}$ for absorption from the electronic ground
state of the oligomer to an excited state with energy $E_j$ is given by \cite{Kn84_73_,FiKnWi91_7880_,MaMo95_14587_}
\begin{equation}
\label{F_nu}
F_{j}=\frac{1-(-1)^{j}}{N+1}\mathrm{ctg}^2\frac{\pi j}{2(N+1)}
\end{equation}
where the oscillator strength of a monomer is taken to be unity.
 From this one sees that the state with $j=1$, which is located at the edge of the exciton band, carries nearly all the oscillator strength. For large $N$ it carries roughly 81\% of the oscillator strength; for  $N<7$ even more than 90\%.
Depending on the sign of $V$ this state is either shifted to lower energies (for $V<0$) or to higher energies (for $V>0$) with respect to the electronic transition energy $\ElTransE$.
Note that with increasing number of monomers $N$ this state is shifted further away from the monomer transition energy, approaching for $N\rightarrow \infty$ the value $E_1=2V$.
Note also that the absorption spectra for $V$ and $-V$ are just mirror images of each other.
This will change drastically when one includes coupling to vibrations.

For the vibrational  environment we consider a continuum of frequencies  so that the
spectral density $J_n(\omega)$ becomes a smooth function. 
Exemplarily we will consider the monomer spectrum shown in
Fig.~\ref{fig:monomer}.
This spectrum is obtained from a spectral density (shown in the inset of Fig.~\ref{fig:monomer}) which is composed of a sum of Lorentzians. 
The zero of energy is located at the transition energy $\ElTransE$.
Due to the strong interaction with the vibrations this spectrum is considerably broadened. It also shows a pronounced asymmetry (it is much steeper at the low energy side). This strong asymmetry is mainly due to coupling to high energy vibrations with vibrational energies around $0.18\, \rm{ eV}$, which is roughly the value of a C=C stretch mode in organic molecules.

 \begin{figure}[bt]
\centering
\psfrag{Monomer}{\hspace{-0.05\mylenunit} \scriptsize Monomer}
\psfrag{N=2}{\hspace{-0.08\mylenunit} \scriptsize $N=2$}
\psfrag{N=3}{\hspace{-0.08\mylenunit} \scriptsize $N=3$}
\psfrag{N=6}{\hspace{-0.08\mylenunit} \scriptsize $N=6$}
\psfrag{N=12}{\hspace{-0.08\mylenunit} \scriptsize $N=12$}
\psfrag{N=21}{\hspace{-0.08\mylenunit} \scriptsize $N=21$}
\psfrag{V=-0.15}{\hspace{-0.08\mylenunit}\scriptsize $V=-0.15$}
\psfrag{V=0.15}{\hspace{-0.08\mylenunit}\scriptsize $V=0.15$}
\psfrag{V=-0.05}{\hspace{-0.08\mylenunit}\scriptsize $V=-0.05$}
\psfrag{V=0.05}{\hspace{-0.08\mylenunit}\scriptsize $V=0.05$}
\psfrag{energy}{\hspace{+0.1\mylenunit}\raisebox{-0.01\mylenunit}{\small Energy $[{\rm eV}]$}}
\psfrag{absorption}{\small\hspace{-0.15\mylenunit}\raisebox{0.01\mylenunit}{Absorption [arb. u.]}}
\psfig{file=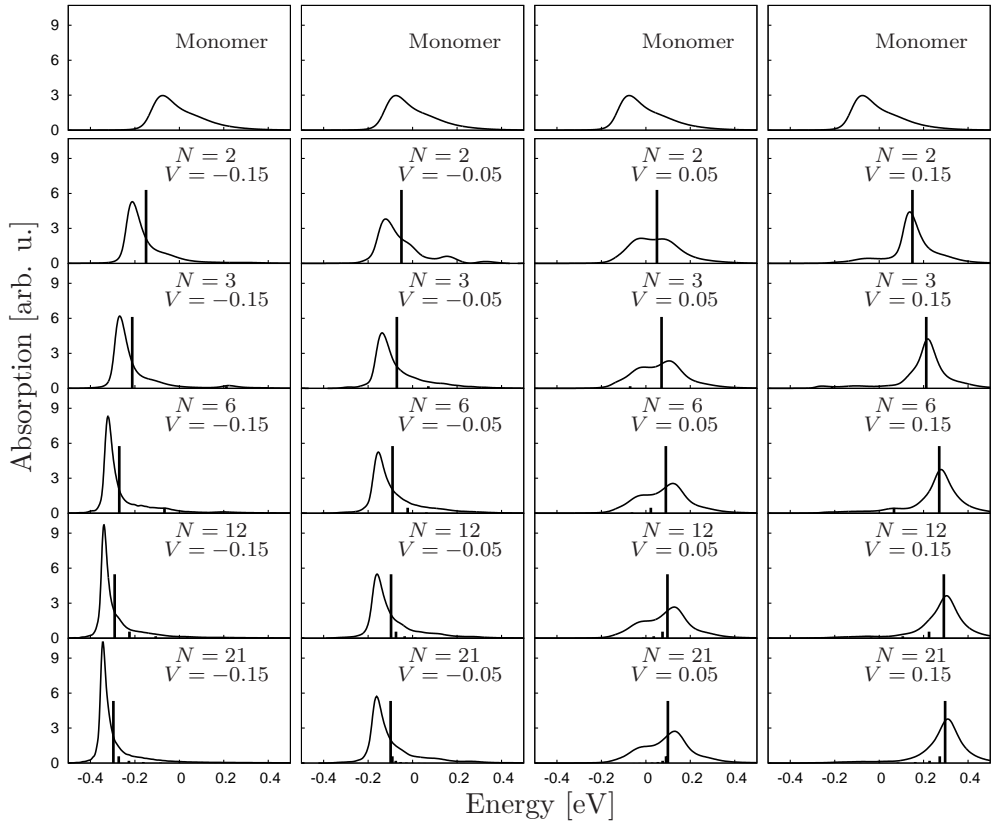,width=5.8in,angle=0}
 \vspace*{8pt}
 \caption{\label{fig:oligomer}Absorption spectra of linear oligomers for different numbers of monomers $N$ and various coupling strength $V$ (in eV). The solid curves are calculated using the method described in section \ref{sec:method} for the spectral density of Fig.\ref{fig:monomer}; sticks are obtaind from the purely electronic Hamiltonian.}
 \end{figure}
We will now take a closer look at the dependence of the oligomer spectra on the number $N$ of monomers  for various values of the interaction strength $V$.
Each column of Fig.~\ref{fig:oligomer} shows, for a distinct value of the dipole-dipole interaction $V$, a sequence of absorption spectra with increasing  length $N$ of the oligomer  from top to bottom. The values of $V$ and $N$ are indicated in the individual plots.
On top of each column the monomer spectrum  from Fig.~\ref{fig:monomer} is shown again for comparison. 
In addition to the spectra calculated including all vibrational modes also ``stick-spectra'' resulting from the purely electronic theory are shown. 

The values chosen for the interaction strength $V$ between neighboring
monomers are $V=\pm 0.15\ \rm{eV}$ (column 1 and 4) and  $V=\pm 0.05\ \rm{eV}$
(column 2 and 3). These values are of the order of magnitude as what is found
in experiment \cite{KoHaKa81_498_,SeWiRe09_13475_}. 

As explained above, the purely electronic stick spectra for opposite sign (but same magnitude) of $V$ are mirror images with respect to each other.
However, as can clearly be seen, this is no longer the case for the spectra including coupling to the vibrations.
Although, in accordance with sum rules \cite{BrHe70_1663_,Ei07_321_,AmVaGr00__}, the mean of each absorption spectrum is centered at the mean of the corresponding purely electronic spectrum, the line-shape is completely different for negative and positive interaction~$V$.

We will now discuss the dependence of the absorption spectra on $N$ and $V$ in more detail.
For large negative interaction $V=-0.15\ \rm{eV}$ (first column) 
 the absorption line-shape considerably narrows upon increasing  the number $N$ of monomers of the oligomer. 
This exemplifies the appearance of the famous so-called J-band \cite{Ko96__}, a very narrow red-shifted peak found in the absorption spectra of large dye-aggregates.
It shows a strong asymmetry, with a steep slope at the low energy side and a long tail extending to higher energies.
Due to this  asymmetry of the line-shape the peak maximum appears shifted to lower energies with respect to the purely electronic ``stick''.
For weaker interaction  $V=-0.05\ \rm{eV}$  (second column) the narrowing  can already be seen, but is less pronounced than for  $V=-0.15\ \rm{eV}$. 
Furthermore, when increasing the number of monomers from 3 to 6 there is little change in the absorption spectrum for $V=-0.05\ \rm{eV}$, while for  $V=-0.15\ \rm{eV}$ there is still a noticeable narrowing. 
In contrast to the narrowing found for negative values of $V$, for the same absolute value $|V|$ but positive $V$, the spectra are much broader and show little resemblance with their negative counterparts.
This is the typical case of the so-called H-band, which is blue-shifted with respect to the monomer absorption. 

Note that for the present values of $V$ there is only minor change in the absorption spectrum for $N>6$, which indicates that for larger $N$ finite size effects will no longer be of great importance.

\section{Summary and Outlook}
\label{sec:conclusions}
In this work we have investigated the properties of one-dimensional linear oligomers using a recently
developed approach based on a stochastic Schr\"odinger equation \cite{RoEiWo09_058301_} to handle  coupling of the electronic excitation to vibrations.
Considering the length dependence of the oligomer spectra as well  as the dependence on the dipole-dipole interaction between the monomers, we found that this approach, although approximate,  is well suited to describe the basic features like the narrow J-band and the broad H-band.
While in this work discussion was restricted to zero temperature, we plan to
study the temperature dependence with the present method and compare to experiment. 

\ack{A.\ E.\ acknowledges discussions  with S.\ Trugman. He also thanks Gerd for the great hospitality in Quito.}

\end{document}